\documentclass[english]{article}
\usepackage[T1]{fontenc}
\usepackage[latin9]{inputenc}
\usepackage{geometry}
\usepackage{color}
\usepackage{babel}
\usepackage{float}
\usepackage{amsmath}
\usepackage{amsthm}
\usepackage{amssymb}
\usepackage{stmaryrd}
\usepackage{graphicx}
\usepackage{setspace}
\usepackage[unicode=true,pdfusetitle,
 bookmarks=true,bookmarksnumbered=false,bookmarksopen=false,
 breaklinks=false,pdfborder={0 0 1},backref=false,colorlinks=false]
 {hyperref}
\begin{document}
\begin{doublespace}
\begin{center}
\textbf{\textcolor{black}{\large{}Solving the Equity Risk Premium
Puzzle}}{\large\par}
\par\end{center}

\begin{center}
\textbf{and}
\par\end{center}

\begin{center}
\textbf{\textcolor{black}{\large{}Inching Towards a Theory of Everything}}{\large\par}
\par\end{center}

\begin{center}
\textbf{Ravi Kashyap}
\par\end{center}

\begin{center}
\textbf{SolBridge International School of Business / City University
of Hong Kong}
\par\end{center}

\begin{center}
\begin{center}
\today
\par\end{center}
\par\end{center}

\begin{center}
\par\end{center}

\begin{center}
\par\end{center}
\end{doublespace}

\begin{center}
\textbf{\textcolor{blue}{\href{https://doi.org/10.3905/jpe.2018.21.2.045}{Edited Version: Kashyap, R. (2018). Solving the Equity Risk Premium Puzzle and Inching Toward a Theory of Everything. The Journal of Private Equity, 21(2), 45-63. }}}\tableofcontents{}\pagebreak{}
\par\end{center}
\begin{doublespace}

\section{Abstract}
\end{doublespace}

\begin{doublespace}
The equity risk premium puzzle is that the return on equities has
far exceeded the average return on short-term risk-free debt and cannot
be explained by conventional representative-agent consumption based
equilibrium models. We review a few attempts done over the years to
explain this anomaly:
\end{doublespace}
\begin{enumerate}
\begin{doublespace}
\item Inclusion of highly unlikely events with low probability (Ugly state
along with Good and Bad), or market crashes, recently also termed
as Black Swans.
\item Slow moving habit, or time-varying subsistence level, added to the
basic power utility function.
\item Allowing for a separation of the inter-temporal elasticity of substitution
and risk aversion, combined with consumption and dividend growth rates
modeled as containing a small persistent expected growth rate component
and a fluctuating volatility which captures time varying economic
uncertainty.
\end{doublespace}
\end{enumerate}
\begin{doublespace}
We explore whether a fusion of the above approaches supplemented with
better methods to handle the below reservations would provide a more
realistic and yet tractable framework to tackle the various conundrums
in the social sciences:
\end{doublespace}
\begin{enumerate}
\begin{doublespace}
\item Unlimited ability of individuals to invest as compared to their ability
to consume.
\item Lack of an objective measuring stick of value which gives rise to
heterogeneous preferences and beliefs.
\item Unintended consequences due to the dynamic nature of social systems,
where changes can be observed and decisions effected by participants
to influence the system.
\item Relaxation of the transversality condition to avoid the formation
of asset price bubbles.
\item How durable is durable? Since nothing lasts forever, accounting for
durable goods to create a comprehensive measure of consumption volatility.
\end{doublespace}
\end{enumerate}
\begin{doublespace}
The world we live in produces fascinating phenomenon despite (or perhaps,
due to) being a hotchpotch of varying doses of the above elements.
The rationale for a unified theory is that beauty can emerge from
chaos since the best test for a stew is its taste. 
\end{doublespace}

Many long standing puzzles seem to have been resolved using different
techniques. The various explanations need to stand the test of time
before acceptance; but then unexpected outcomes set in and new puzzles
emerge. As real analysis and limits tell us: we are getting Closer
and Closer; Yet it seems we are still Far Far Away...
\begin{doublespace}

\section{Notation and Terminology for Key Results}
\end{doublespace}

Unless explicitly specified or respecified for a section, all symbols
apply throughout the entire paper. Please consult corresponding sections
including the appendix for more details.

\subsection{Equity Premium Puzzle}
\begin{itemize}
\begin{doublespace}
\item $R_{t}^{i},R_{t}^{e},R_{t}^{f},R_{t}^{mv}$, are the returns on any
security $i$, equity, risk free security and any portfolio on the
mean variance frontier at time $t$. We use smaller case to denote
the natural logarithm of the corresponding returns, $r^{f}=\ln R^{f}$
and so on.
\item $m$, is the discount factor.
\end{doublespace}
\item $E\left(X\right);\text{Var}\left(X\right);\sigma\left(X\right)$,
are the mean, variance and standard deviation of random variable $X$.
$\text{Cov}\left(X,Y\right)\;;\;\rho_{X,Y}$ are the covariance and
the correlation between random variables $X$ and $Y$.
\begin{doublespace}
\item $\left(\alpha,\beta\right)$, are the risk aversion coefficient and
the subjective time discount factor. $\delta=-\ln\beta$ is also the
subjective discount factor.
\end{doublespace}
\end{itemize}

\subsubsection{Hansen and Jagannathan Bound in Appendix \ref{subsec:Hansen-and-Jagannathan}}
\begin{itemize}
\item $\alpha$ also measures the curvature of the utility function; to
start with, we assume a power utility function of the constant relative
risk aversion class of the form, $U\left(C_{t},\alpha\right)=\frac{C_{t}^{1-\alpha}-1}{1-\alpha}$.
\end{itemize}

\subsubsection{Consumption Growth and Interest Rates in Appendix \ref{subsec:Consumption-Growth-and}}
\begin{itemize}
\begin{doublespace}
\item $C_{t}$ is the consumption at time $t$. We also set, $\Delta c_{t+1}\equiv\ln C_{t+1}-\ln C_{t}$.
\end{doublespace}
\end{itemize}

\subsubsection{Mehra and Prescott variation with Markov Process in Appendix \ref{subsec:Mehra-and-Prescott}}
\begin{itemize}
\begin{doublespace}
\item $\pi\in R^{n}$, is the vector of stationary probabilities for the
ergodic Markov process governing consumption growth.
\item $p^{e}\;;\;p^{f}$ , are the prices of the equity and risk free securities.
\end{doublespace}
\item $y_{t}$ is the firm's dividend payment in the period $t$. The firm's
output is constrained to be less than or equal to $y_{t}$.
\item $x_{t+1}\in\left\{ \lambda_{1},\ldots,\lambda_{n}\right\} $ is the
growth rate of the dividend payment $y_{t}$.
\begin{doublespace}
\item $\left\{ \phi_{ji}\right\} $, is the transition probability between
states $i$ and $j$.
\end{doublespace}
\item $\left(\alpha,\beta\right)$and $\left(\mu,\phi,\gamma\right)$ are
parameters that define preferences and technology respectively.
\end{itemize}

\subsection{Highly Unlikely Events}
\begin{itemize}
\begin{doublespace}
\item $\eta$, is the low crash probability.
\end{doublespace}
\item $\psi$ is a fraction or a combination of the other parameters such
that $\lambda_{1}>\lambda_{2}>\lambda_{3}$.
\end{itemize}

\subsection{Force of Habit}
\begin{itemize}
\begin{doublespace}
\item $X_{t}\;;\;S_{t}$, denote the level of habits and the surplus consumption
ratio included in the utility function. Also, $s_{t+1}\equiv\ln S_{t+1}$.
\end{doublespace}
\item $\lambda\left(s_{t}\right)$ , is the sensitivity function.
\item $\eta_{t}$, is the local curvature when the utility function is modified
to include the level of habits and the surplus consumption ratio.
\item $\varrho$, $g$, $\bar{s}$ are parameters defined in the heteroskedastic
AR(1) process for the log surplus consumption ratio.
\end{itemize}
\begin{doublespace}

\subsection{Long Run Risks and Survivors}
\end{doublespace}
\begin{itemize}
\item $\varphi\geq0$ is the parameter for Inter-temporal Elasticity of
Substitution (IES). 
\item $G_{t+1}$ is the aggregate growth rate of consumption.
\item $R_{i,t+1}$ is the gross return on asset $i$.
\item $R_{a,t+1}$ is the unobservable gross return on an asset that delivers
aggregate consumption as its dividend each period. 
\item $R_{m,t+1}$ is the observable return on the market portfolio and
the return on the aggregate dividend claim. 
\item $q_{t}$ is a small persistent predictable component in the consumption
and dividend growth rates. 
\item $g_{t+1}$ and $g_{d,t+1}$ are the growth rates on consumption and
dividends. 
\item $\sigma_{t+1}$, represents the time-varying economic uncertainty
incorporated in consumption growth rate and $\sigma^{2}$ is its unconditional
mean.
\item $r_{f,t}$ and $E_{t}\left(r_{m,t+1}-r_{f,t}\right)$ are the risk
free rate and the equity premium in the presence of time-varying economic
uncertainty.
\item $\beta_{m,e}\:,\:\lambda_{m,e}\:,\:\beta_{m,w}\:,\:\lambda_{m,w}$
are combinations of other parameters.
\end{itemize}
\begin{doublespace}

\subsection{Heterogeneous Agents}
\end{doublespace}
\begin{itemize}
\item $C_{it+1}$ is individual consumption growth, determined by an independent
idiosyncratic shock $\eta_{it}$, such that, $\ln\left(\frac{C_{it+1}}{C_{it}}\right)=\eta_{it+1}b_{t+1}-\frac{b_{t+1}^{2}}{2}\quad;\eta_{it}\sim N\left(0,1\right)$
.
\item $b_{t+1}$ is the cross-sectional standard deviation of consumption
growth. 
\end{itemize}
\begin{doublespace}

\section{Equity Premium Puzzle}
\end{doublespace}

\begin{doublespace}
The equity risk premium puzzle is that the return on equities has
far exceeded the average return on short-term risk-free debt and cannot
be explained by conventional representative-agent consumption based
equilibrium models.

(Mehra and Prescott 1985) study a class of competitive pure exchange
economies for which the equilibrium growth rate process on consumption
and equilibrium asset returns are stationary. Attention is restricted
to economies for which the elasticity of substitution for the composite
consumption good between the year $t$ and year $t+1$ is consistent
with findings in micro, macro and international economics. In addition,
the economies are constructed to display equilibrium consumption growth
rates with the same mean, variance and serial correlation as those
observed for the U.S. economy in the 1889-1978 period. They find that
for such economies, the average real annual yield on equity is a maximum
of four-tenths of a percent higher than that on short-term debt, in
sharp contrast to the six percent premium observed. Their results
are robust to non-stationarities in the means and variances of the
economies growth processes.

Historically the average return on equity has far exceeded the average
return on short-term virtually default-free debt. Over the ninety-year
period 1889-1978 the average real annual yield on the Standard and
Poor 500 Index was seven percent, while the average yield on short-term
debt was less than one percent. They address the question whether
this large differential in average yields can be accounted for by
models that abstract from transactions costs, liquidity constraints
and other frictions absent in the Arrow-Debreu set-up (Arrow and Debreu
1954; McKenzie 1954, 1959; Debreu 1987). They conclude that, for the
class of economies considered, most likely some equilibrium model
with a friction will be the one that successfully accounts for the
large average equity premium.

The below bound due to (Hansen and Jagannathan 1991) can be used to
understand the equity premium puzzle. Appendix \ref{subsec:Hansen-and-Jagannathan}
has the steps to arrive at this result.
\begin{equation}
\left|\frac{E\left(R^{mv}\right)-R^{f}}{\sigma\left(R^{mv}\right)}\right|\approx\alpha\sigma\left(\Delta c_{t+1}\right)
\end{equation}

\end{doublespace}

The excess return on equity instruments has been in the seven to nine
percent range. The returns just after the second world war are around
9\% with a standard deviation of about 16\%. The risk free rate has
been stable around the 1\% level. Aggregate non-durable and services
consumption growth had a mean and standard deviation of 1\%. To explain
these observed results, the risk aversion coefficient, $\alpha$,
needs to be around 50. If we consider the actual correlation between
annual returns and non-durables plus services consumption growth,
which is around 0.2, $\alpha$, needs to be around 250.

\begin{doublespace}
(Mehra and Prescott 1985) restrict the value of $\alpha$ to be a
maximum of ten based on evidence from other studies, ``The parameter
$\alpha$, which measures people's willingness to substitute consumption
between successive yearly time periods is an important one in many
fields of economics. (Arrow 1971) summarizes a number of studies and
concludes that relative risk aversion with respect to wealth is almost
constant. He further argues on theoretical grounds that $\alpha$
should be approximately one. (Friend and Blume 1975) present evidence
based upon the portfolio holdings of individuals that $\alpha$ is
larger, with their estimates being in the range of two. (Kydland and
Prescott 1982), in their study of aggregate fluctuations, found that
they needed a value between one and two to mimic the observed relative
variability of consumption and investment. (Altug 1983), using a closely
related model and formal econometric techniques, estimates the parameter
to be near zero. (Kehoe and Richardson 1984), studying the response
of small countries balance of trade to terms of trade shocks, obtained
estimates near one, the value posited by Arrow. (Hildreth and Knowles
1982) in their study of the behavior of farmers also obtain estimates
between one and two. (Tobin and Dolde 1971), studying life cycle savings
behavior with borrowing constraints, use a value of 1.5 to fit the
observed life cycle savings patterns.''

Looking at this from another angle we get another set of inconsistencies.
A high value of risk aversion, $\alpha=$ 50 to 250 implies a very
high risk free rate of 50-250\% as seen from the below relation between
consumption growth and interest rates (Appendix \ref{subsec:Consumption-Growth-and}
has the steps).
\begin{equation}
r^{f}=\ln R^{f}=\delta+\alpha E\left(\Delta c_{t+1}\right)-\frac{\alpha^{2}}{2}\sigma^{2}\left(\Delta c_{t+1}\right)
\end{equation}
To get a reasonable interest rate (usually around 1\%), we need a
subjective discount factor of $\delta=$ -0.5 to -2.5 or -50\% to
-250\% (or, $\beta=e^{-\delta}>1$), which seems unreasonable since
people prefer earlier utility. 

(Mehra and Prescott 1985) start with a pure exchange model, (Lucas
1978) and include a variation (Mehra 1988) such that the growth rate
of consumption follows a Markov process in contrast to the Lucas tree
economy where the consumption level follows a Markov process. There
is one productive unit or firm producing the perishable consumption
good and there is one equity share that is competitively traded. Since
only one productive unit is considered, the return on this share of
equity of equity is also the return on the market. 
\end{doublespace}

With two states, the Markov process growth rates and transition probabilities
are (see Appendix \ref{subsec:Mehra-and-Prescott} for the steps),
\begin{equation}
\lambda_{1}=1+\mu+\gamma,\quad\lambda_{2}=1+\mu-\gamma
\end{equation}
\begin{equation}
\phi_{11}=\phi_{22}=\phi,\quad\phi_{12}=\phi_{21}=1-\phi
\end{equation}
The above parameters $\left\llbracket \left(\alpha,\beta\right)\right.$
and $\left\{ \left(\mu,\phi,\gamma\right)\right.$ $\equiv$ elements
of $\left[\phi_{ij}\right]$ and $\left.\left.\left[\lambda_{i}\right]\right\} \right\rrbracket $,
define preferences and technology respectively; they are estimated
using method of moments by matching the mean, variance and first order
auto-correlation of the growth rate of per-capita consumption. Based
on the estimated parameters, the maximum value of the equity premium
is 0.35 percent.
\begin{doublespace}

\section{Deep Dive into Possible Explanations\label{sec:Deep-Dive-into}}
\end{doublespace}
\begin{doublespace}

\subsection{Highly Unlikely Events}
\end{doublespace}

\begin{doublespace}
(Rietz 1988) is one of the, if not the, earliest attempt to resolve
the equity premium puzzle. Their departure from the Mehra Prescott
specification is mainly in the assumption of three possible growth
rates in three states of the world. We term them the good, bad and
ugly states. The good and bad states are the same as before, but when
things get really bad or the market crashes, we end up in the ugly
state or in a depression like episode. Equity returns vary little
from the norm in good and bad times, but there are rare or low probability
events or crashes when consumption falls drastically and equity returns
are far below average. It is worth noting that in their original paper,
Mehra and Prescott consider a four state Markov process (though the
growth rates can only be either good, poor or average), the probability
of the states are not significantly different, with average times
twice as likely as either poor or good, and the maximum premium explained
in this case is only 0.39 percent.

The growth rates in the three states and the transition probability
matrix with a disaster scenario are,
\begin{equation}
\lambda_{1}=1+\mu+\gamma,\quad\lambda_{2}=1+\mu-\gamma,\quad\lambda_{3}=\psi\left(1+\mu\right)
\end{equation}
\begin{equation}
\Phi=\left[\begin{array}{ccc}
\phi & 1-\phi-\eta & \eta\\
1-\phi-\eta & \phi & \eta\\
\frac{1}{2} & \frac{1}{2} & 0
\end{array}\right]
\end{equation}
Here, $\psi$ is a fraction or a combination of the other parameters
such that $\lambda_{1}>\lambda_{2}>\lambda_{3}$. A crash will only
follow state 1 or 2 and never occurs twice in a row and happens with
the low crash probability, $\eta$. The estimation of the parameters
proceeds in a similar fashion as (Mehra and Prescott 1985) for given
crash probabilities; (Reitz 1988) takes $\eta\in\left[0.0001,0.2\right]$;
$\alpha\in\left(0\right.,\left.10\right]$; $\beta\in\left(0,1\right)$
and the equity premium is explained successfully with this setup.
(Mehra and Prescott 1988) comment on the validity of the this unlikely
state assumption, especially about the value of the risk aversion
parameter, $\alpha$, being set to 10 and the plausibility of the
occurrence of real disaster scenarios. 
\end{doublespace}

(Barro 2006) acknowledges that the work of Rietz has been under appreciated
and calibrates disaster probabilities, especially the sharp contractions
associated with the the two world wars and the great depression. His
empirical analysis measures the size and frequency of economic disasters
and estimates a disaster probability of 1.5 to 2 percent per year
with a distribution of declines in per-capita GDP ranging between
15 percent to 64 percent. He also points out that considering capital
formation; the duration of disasters; disaster scenarios and real
estate prices; and depreciation and creation of Lucas trees, would
be useful extensions. 

The recent work of (Taleb 2007) and the occurrence of Black Swans
(which are unexpected events and cannot be probabilistically anticipated
but they still happen catching us off guard) is a also fillip in this
line of thinking. (Weitzman 2007) offers a single unified theory for
the equity premium puzzle and two other asset return puzzles: risk-free
rate and equity-volatility puzzles, based on the idea that what is
learn-able about the future stochastic consumption-growth process
from any number of past empirical observations must fall far short
of full structural knowledge. 

The equity-volatility puzzle refers to the empirical fact that actual
returns on a representative stock market index have a variance some
two orders of magnitude larger than the variance of any consumption-dividend-like
fundamental in the real economy that might possibly be driving them
or that might be relevant for welfare calibration. The risk-free-rate
puzzle refers to the five percentage points or so discrepancy between
the interest rate that is predicted by any standard rational expectations
equilibrium (REE) formula and what is actually observed. The key characteristic
of REE (defining it as a proper subset of the set of all rationally
formed Bayesian equilibria) is the imposed extra assumption that the
subjective probability distribution of outcomes believed by agents
within an economic system equals the objective frequency distribution
actually generated by the system itself. It is effectively a dynamic
stochastic general equilibrium where all reduced-form structural parameters
of the data-generating process are known, presumably because they
have already been learned previously as some kind of an ergodic limit
from a sufficiently large sample.
\begin{doublespace}

\subsection{Force of Habit}
\end{doublespace}

\begin{doublespace}
The predictability of returns from prices and dividends can be explained
based on the idea that people get less risk averse when wealth or
consumption increases and more risk averse as wealth decreases. The
level of consumption and wealth increases over time, while the equity
premia have not declined. So tying risk aversion to the level of consumption
or wealth, relative to a historical trend or some previous benchmark,
might hold some potential solutions.
\end{doublespace}

Marshall (1920), probably one of the earliest works on this topic,
discussed the notion that tastes can be cultivated and that they are
affected by past consumption. (Duesenberry 1949) proposed an individual
consumption function that depended on the current consumption of other
people. Also known as the relative income hypothesis, this was an
attempt to rationalize the well established differences between cross-sectional
and time-series properties of consumption data. (Pollak 1970) considers
dynamic utility functions by allowing some or all of its parameters
to depend on past consumption and notes that the dominant assumption
upto that point in time was that an individual's utility function
depends on his own consumption, but not on the consumption of others.
Hence, by allowing some or all of the parameters of an individual's
utility function to depend on the consumption of others, interdependence
can be incorporated into the theory of consumer behavior. 

(Constantinides 1990) is an important theoretical work on the subject
of habit formation. (Campbell and Cochrane 1999) specify that people
slowly develop habits for higher or lower consumption. We have an
endowment economy with i.i.d. consumption growth. We modify the utility
function to include a term for the level of habits, $X_{t}$. The
sensitivity function, $\lambda\left(s_{t}\right)$ is specified to
satisfy three conditions, 1) Risk free rate is constants; 2) habit
is predetermined at the steady state $\left(s_{t}-\bar{s}\right)$;
and 3) habit moves negatively with consumption everywhere or equivalently,
habit is predetermined near the steady state. Local curvature, $\eta_{t}$,
depends on how far consumption is above the habit, as well as $\alpha.$

As consumption falls toward habit, people become much less willing
to tolerate further falls in consumption; they become very risk averse.
Thus, a low power coefficient $\alpha$ can still mean a high and
time-varying curvature. High curvature means that the model can explain
the equity premium, and curvature that varies over time as consumption
rises in booms and falls toward habit in recessions, means that the
model can explain a time-varying and counter-cyclical (high in recessions,
low in booms) Sharpe ratio, despite constant consumption volatility
$\sigma_{t}\left(\Delta c\right)$ and correlation $\text{corr}\left(\Delta c,r\right)$.
But higher curvature implies high and time-varying interest rates.
This model gets around interest rate problems with precautionary saving.
Suppose we are in a bad time, in which consumption is low relative
to habit. People want to borrow against future higher consumption,
and this force should drive up interest rates. (Habit models tend
to have very volatile interest rates.) However, people are also much
more risk averse when consumption is low. This consideration induces
them to save more, in order to build up assets against the event that
tomorrow might be even worse. This precautionary desire to save drives
down interest rates. The sensitivity function specification, $\lambda\left(s_{t}\right)$,
makes these two forces exactly offset, leading to constant real rates
(see Appendix \ref{subsec:Force-of-Habit} for the steps to derive
the below result). 
\begin{equation}
\left|\frac{E_{t}\left(R_{t+1}^{e}\right)}{\sigma_{t}\left(R_{t+1}^{e}\right)}\right|=\sqrt{\left\{ e^{\alpha^{2}\sigma^{2}\left[1+\lambda\left(s_{t}\right)\right]^{2}}-1\right\} }\approx\alpha\sigma\left[1+\lambda\left(s_{t}\right)\right]
\end{equation}

\begin{doublespace}
The mean and standard deviation of log consumption growth are set
to match consumption data. The serial correlation parameter $\varrho$
is chosen to match the serial correlation of the logarithm of the
ratio of price divided by dividend. $\beta$ or the subjective discount
factor is chosen to match the risk free rate with the average return
on real treasury bills. The risk aversion parameter $\alpha$ is then
searched so that the returns on the consumption claim matches the
ratio of the unconditional mean and unconditional standard deviation
of excess returns. 

In contrast to the Reitz model of a small probability of a very large
negative consumption shock, investors fear stocks because they do
badly in occasional serious recessions unrelated to the risks of long-run
average consumption growth.
\end{doublespace}
\begin{doublespace}

\subsection{Long Run Risks and Survivors}
\end{doublespace}

\begin{doublespace}
(Bansal and Yaron 2004) allow for a separation of the inter-temporal
elasticity of substitution (IES) and risk aversion, combined with
consumption and dividend growth rates modeled as containing a small
persistent expected growth rate component and a fluctuating volatility
which captures time varying economic uncertainty. (Epstein and Zin
1989; Weil 1989; Weil 1990; Chen, Favilukis and Ludvigson 2013) present
estimates of key preference parameters in the recursive utility model,
evaluate the model\textquoteright s ability to fit asset return data
relative to other asset pricing models and investigate the implications
of such estimates for the unobservable aggregate wealth return. Building
on the recursive or non-expected utility preferences, which distinguish
attitudes toward risk from behavior toward inter-temporal substitution,
we can arrive at the following expressions for the risk free rate,
$r_{f,t}$, and the equity premium, $E_{t}\left(r_{m,t+1}-r_{f,t}\right)$,
in the presence of time-varying economic uncertainty (Appendix \ref{subsec:Long-Run-Risks}
has more details),
\begin{equation}
r_{f,t}=-\theta\ln\beta+\frac{\theta}{\varphi}E_{t}\left[g_{t+1}\right]+\left(1-\theta\right)E_{t}\left[r_{a,t+1}\right]-\frac{1}{2}\text{var}_{t}\left[\frac{\theta}{\varphi}g_{t+1}+\left(1-\theta\right)r_{a,t+1}\right]
\end{equation}
\begin{equation}
E_{t}\left(r_{m,t+1}-r_{f,t}\right)=\beta_{m,e}\lambda_{m,e}\sigma_{t}^{2}+\beta_{m,w}\lambda_{m,w}\sigma_{w}^{2}-0.5\text{var}_{t}\left(r_{m,t+1}\right)
\end{equation}

\end{doublespace}

$\varphi\geq0$ is the IES parameter. $G_{t+1}$ is the aggregate
growth rate of consumption; $R_{a,t+1}$ is the unobservable gross
return on an asset that delivers aggregate consumption as its dividend
each period; $R_{m,t+1}$ is the observable return on the market portfolio
and the return on the aggregate dividend claim; $g_{t+1},r_{a,t+1},r_{m,t+1}$are
the logarithms of the variables just mentioned. $q_{t}$ is a small
persistent predictable component in the consumption and dividend growth
rates. The growth rates on consumption $g_{t+1}$ and dividends $g_{d,t+1}$
are modeled as shown in appendix \ref{subsec:Long-Run-Risks}, (following
Campbell and Shiller 1988, who use similar log linear approximations
to show that price-dividend ratios seem to predict long-horizon equity
returns), $\sigma_{t+1}$, represents the time-varying economic uncertainty
incorporated in consumption growth rate and $\sigma^{2}$ is its unconditional
mean. There is an assumption that the shocks are uncorrelated, and
allow for only one source of economic uncertainty to affect consumption
and dividends. $\beta_{m,e}\:,\:\lambda_{m,e}\:,\:\beta_{m,w}\:,\:\lambda_{m,w}$
are combinations of other parameters.

A simpler specification can set $g_{t+1}=\mu+q_{t}+\sigma\eta_{t+1}$.
But since the economic uncertainty, $\sigma$, is constant, the conditional
risk premium and the conditional volatility of the market portfolio
is constant and hence their ratio, the Sharpe ratio is also constant.
The long run risk or time varying uncertainty gives a large value
for the equity premium while the separation between the IES parameter
and risk aversion ensures that the risk free rate remains small. 

(Bansal, Kiku and Yaron 2010) is a generalized long run risks model
incorporating a cyclical component in aggregate consumption and dividends
and Poisson jumps in the processes. (Drechsler and Yaron 2011) demonstrate
conditions under which the variance premium, defined as the difference
between the squared Chicago Board Options Excange volatility index
(VIX) and the expected realized variance, displays significant time
variation and return predictability. They show that a calibrated,
generalized long-run risks model generates a variance premium with
time variation and return predictability that is consistent with the
data, while simultaneously matching the levels and volatilities of
the market return and risk-free rate. Using book-to-market, momentum,
and size-sorted portfolios, (Bansal, Dittmar and Lundblad 2005) show
that economic risks in cash flows, measured via the cash flow beta
(larger cash flow beta implies higher aggregate consumption risk),
can account for a significant portion of differences in risk premia
across assets. (Jagannathan and Marakani 2015) show that the dependence
of several asset pricing models on long-run risks, implies that the
state of the economy can be captured by factors derived from the price-dividend
ratios of stock portfolios. They relate the Fama-French model and
the Bansal-Yaron and Merton inter-temporal asset pricing models by
using two factors with small growth and large value minus small growth
tilts. {[}Fama and French (1993) interpret the SMB (Small Minus Big)
and HML (High Minus Low) factors, constructed from six size / book-to-market
portfolios, as innovations to state variables in the Merton (1973)
inter-temporal capital asset pricing model; these papers are among
the foundations of asset pricing{]}.

\subsubsection{Market Survivor Bias}

Another thread of explanation is based on the market survivor bias
argument of (Brown, Goetzmann, and Ross, 1995). Empirical analysis
of rates of return, implicitly condition on the security surviving
into the sample. Such conditioning can induce a spurious relationship
between observed return and total risk for those securities that survive
to be included in the sample. The average return for a market that
survives many potentially cataclysmic challenges is likely to be higher
than the expected return. This suggests that past average growth rates
are, if anything, upward biased estimates of future growth. (Fama
and French 2002) estimate the equity premium using dividend and earnings
growth rates to measure the expected rate of capital gain. Their estimates
for 1951 to 2000, 2.55 percent and 4.32 percent, are much lower than
the equity premium produced by the average stock return, 7.43 percent,
suggesting that the high average return for 1951 to 2000 is due to
a decline in discount rates that produces a large unexpected capital
gain. They conclude that the average stock return of the last half-century
is a lot higher than expected.
\begin{doublespace}

\section{Possibilities for a Deeper Dive into a Theory of Everything }
\end{doublespace}

Each of the elegant solutions considered thus far can claim some success
in explaining the equity puzzle. All these models are an artifact
of having many parameters in the model, so that some parameters can
be calibrated to explain particular facets of a phenomenon and other
parameters can be set to explain related but different facets of the
same phenomenon. We need to be wary that the forces in each solution
are not acting in isolation: unlikely events are likely to happen
independent of how consumers modify their behavior or develop any
longer term habits, and long run risks would still persist (though
some dependencies cannot be ruled out, for example, consumption habits
could change based on unlikely events, but they could also change
based on individual health or lifestyle choices). Hence, a consistent
and complete theory needs to combine elements of all the above solutions
and also be able to explain a few other fundamental observations.

\begin{doublespace}
As a first step, we recognize that one possible categorization of
different fields can be done by the set of questions a particular
field attempts to answer. The answers to the questions posed by any
field can come from anywhere or from phenomenon studied under a combination
of many other fields. Hence, we need to keep in mind that the answers
to the questions posed under the realm of economics can come from
diverse fields such as physics, biology, mathematics, chemistry, and
so on.

Hence, before we consider the scope and components of a Theory of
Everything for Economics, let us review similar attempts that have
been going on for many decades in physics. The ``Theory of Everything''
(TOE) is a term for the ultimate theory of the universe (Tegmark 1998;
Laughlin and Pines 2000), a set of equations capable of describing
all phenomena that have been observed, or that will ever be observed.
This would be an all-embracing and self-consistent physical theory
that summarizes everything that there is to know about the workings
of the physical world. (Tegmark 1998; 2008) We can divide TOEs into
two categories depending on their answer to the following question:
Is the physical world purely mathematical, or is mathematics merely
a useful tool that approximately describes certain aspects of the
physical world? More formally, is the physical world isomorphic to
some mathematical structure?

(Cao, Cao and Qiang 2015) discuss why none of the existing theories
(the Theory of Relativity, the Big-Bang, or the Standard Model) can
truly serve as the foundation of physics, because they cannot answer
the fundamental questions, such as why positive and negative charges
exist, why quantum numbers exist, and why electron has mass and never
decays. They admit that, the fundamental questions are ignored because
they are simply too hard to answer. (Cao and Cao 2013) propose a framework,
``Unified Field Theory'', that attempts to provide a real foundation
and to answer the fundamental questions by starting from space-time-energy-force,
the common root for everything, conceptual or physical, to explain
and predict the motion, interaction and configuration of matter. (Barrow
1991, 2007) are excellent discussions on the essential components
that a successful theory of everything should possess and recent research
into the quest for this holy grail.

To find such a common root in the social sciences, we use an existing
definition of economics, which calls it the social science of satisfying
unlimited wants with limited resources (End-note \ref{enu:Unlimited Limited}).
This omnipresent and omnipotent scarcity implies that agents will
endeavor to get more from less. Coupling this fundamental motivation
with the lack of an objective measuring stick of value, leads to an
exchange or a trade (perhaps, only a trade-off sometimes); which is
one of the cornerstones of economics. A trade requires a decision
and it is common to estimate the future value of the item to be traded
or a prediction is made to guide this decision.
\begin{equation}
\left(\text{Trying to get More from Less}\right)\quad\&\quad\left(\text{Difference in Assessment of Value}\right)\Longrightarrow\left(\text{The Need for a Trade}\right)
\end{equation}
\begin{equation}
\left(\text{Prediction}\right)\quad\&\quad\left(\text{Decision}\right)\Longleftrightarrow\left(\text{Trade}\right)
\end{equation}
We can then draw the following parallel (Table \ref{tab:Roots-of-Physics})
to the common roots in the physical world and in the social sciences,
\end{doublespace}

\begin{table}[H]
\begin{doublespace}
\[
\begin{array}{cc}
\underline{\mathbf{Physics}} & \underline{\mathbf{Economics}}\\
Space & Scarcity\\
Time & Subjectivity\\
Energy & Predictions\\
Force & Decisions
\end{array}
\]

\end{doublespace}

\caption{Roots of Physics and Economics\label{tab:Roots-of-Physics}}
\end{table}
The elements we discuss can be categorized into these buckets, though
it should be clear that these prongs are overlapping,
\begin{enumerate}
\begin{doublespace}
\item Scarcity

\end{doublespace}\begin{enumerate}
\begin{doublespace}
\item Force of habit
\item How Durable is Durable?
\end{doublespace}
\end{enumerate}
\begin{doublespace}
\item Subjectivity

\end{doublespace}\begin{enumerate}
\begin{doublespace}
\item Consumption versus investment ability
\end{doublespace}
\item Heterogeneous agents
\end{enumerate}
\begin{doublespace}
\item Predictions

\end{doublespace}\begin{enumerate}
\begin{doublespace}
\item Highly unlikely events
\item Long run risks
\end{doublespace}
\end{enumerate}
\begin{doublespace}
\item Decisions

\end{doublespace}\begin{enumerate}
\begin{doublespace}
\item Unintended consequences
\item Transversality condition
\end{doublespace}
\end{enumerate}
\end{enumerate}
We need to explore further whether a fusion of the solutions discussed
in section \ref{sec:Deep-Dive-into} supplemented with better methods
to handle the below reservations would provide a more realistic and
yet tractable framework to tackle the various conundrums in the social
sciences. The world we live in produces fascinating phenomenon despite
(or perhaps, due to) being a hotchpotch of varying doses of all these
elements. The rationale for a unified theory is that beauty can emerge
from chaos since the best test for a stew is its taste.
\begin{doublespace}

\subsection{Consumption versus Investment Ability}
\end{doublespace}

\begin{doublespace}
Despite the several advances in the social sciences and in particular
economic and financial theory, we have yet to discover an objective
measuring stick of value, a so called, True Value Theory. While some
would compare the search for such a theory, to the medieval alchemist's
obsession with turning everything into gold, for our present purposes,
the lack of such an objective measure means that the difference in
value as assessed by different participants can effect a transfer
of wealth. This forms the core principle that governs all commerce
that is not for immediate consumption in general, and also applies
specifically to all investment related traffic, which forms a great
portion of the financial services industry and hence the mainstay
of asset pricing. (Kashyap 2014) looks at the use of a feedback loop
to aid in the market making of financial instruments. 
\end{doublespace}

Although, some of this is true for consumption assets; because the
consumption ability of individuals and organizations is limited and
their investment ability is not, the lack of an objective measure
of value affects investment assets in a greater way and hence investment
assets and related transactions form a much greater proportion of
the financial services industry. Consumption assets do not get bought
and sold, to an inordinate extent, due to fluctuating prices, whereas
investment assets will (Hull 2010 has a description of consumption
and investment assets, specific to the price determination of futures
and forwards; Kashyap 2014 has a more general discussion). 

\begin{figure}[H]
\includegraphics[width=15cm]{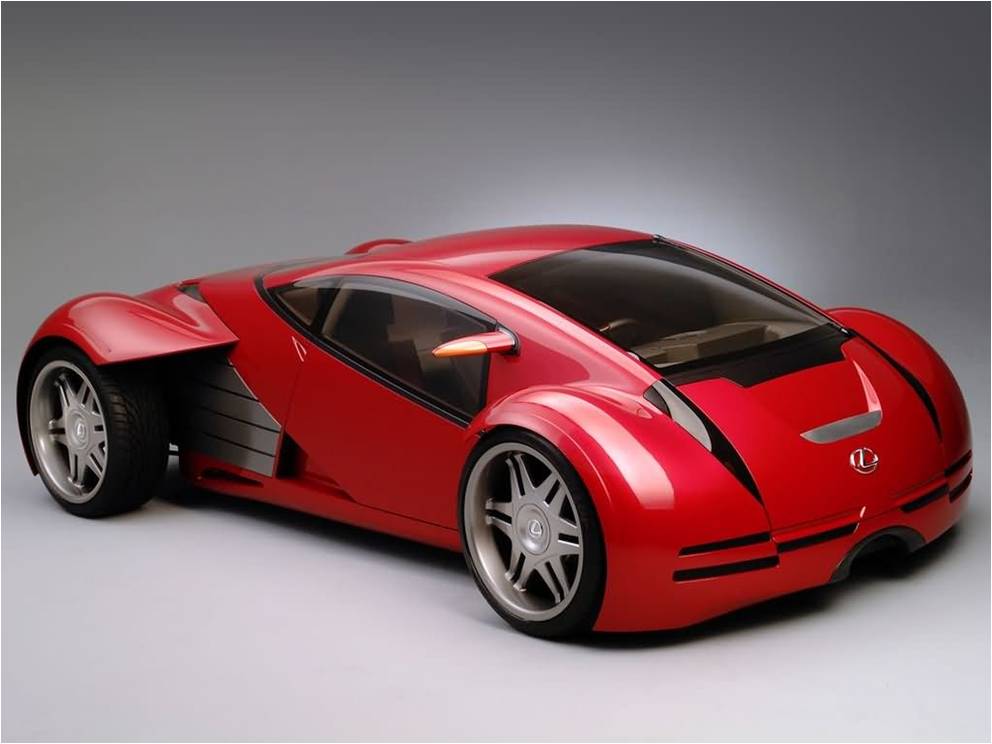}

\caption{Value of Car\label{fig:Value-of-Car}}

\end{figure}

\begin{figure}[H]
\includegraphics[width=15cm]{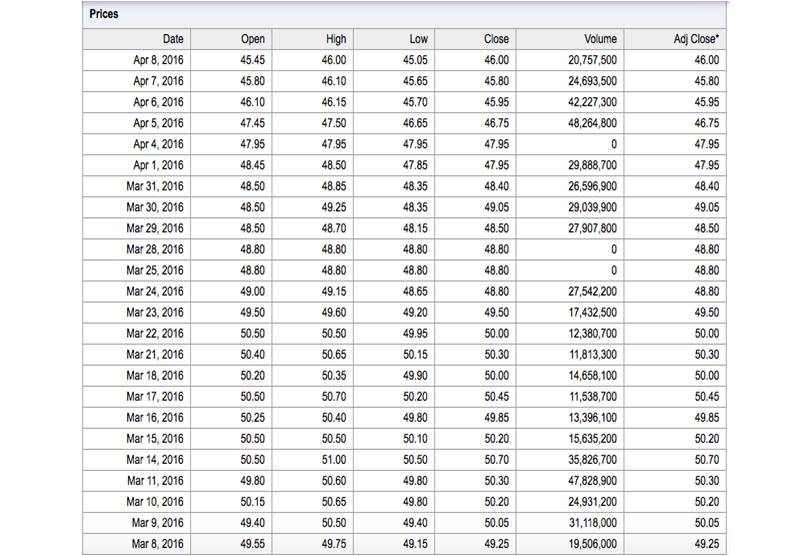}

\caption{Time Series of Close Prices\label{fig:Time-Series-of}}
\end{figure}

We can pose two questions based on figures \ref{fig:Value-of-Car}
and \ref{fig:Time-Series-of}.
\begin{enumerate}
\item What is the value of this car in USD? 
\item What will be the next closing price for this time series of security
prices, assuming we are on the last date of the time series?
\end{enumerate}
From the different answers that different people come up with (and
also from the different questions that people ask in order to answer
these two questions), it should be evident that most measures of value
are subjective. It should also be clear that the price of the security
(investment asset) fluctuates more than the price of the car (consumption
asset) even after a value is agreed upon. We need to devise appropriate
measures to capture how big consumption ability is when compared to
investment ability. In essence, what we are comparing is the relative
size or the cardinality of two infinite sets, a routine question from
real analysis. (Courant and Robbins 1996; Rudin 1964; Royden and Fitzpatrick
1988)

\begin{doublespace}

\end{doublespace}
\begin{doublespace}

\subsection{Heterogeneous Agents}
\end{doublespace}

\begin{doublespace}
The lack of an objective measuring stick of value also gives rise
to heterogeneous preferences and beliefs. (Constantinides and Duffie
1996) provide a clever and simple model with relatively standard preferences,
in which idiosyncratic risk can be tailored to generate any pattern
of aggregate consumption and asset prices. Idiosyncratic risk stories
face two severe challenges. First, the basic pricing equation applies
to each individual. If we are to have low risk aversion and power
utility, the required huge volatility of consumption is implausible
for any individual. Second, if you add idiosyncratic risk uncorrelated
with asset returns, it has no effect on pricing implications. \{(Constantinides
1982) looks at other issues that come up with consumer heterogeneity\}.
Say, agent A gets more income when the market is high, and agent B
gets more income when it is low. But then A will short the market,
B will go long, and they will trade away any component of the shock
that is correlated with the returns on available assets. Shocks uncorrelated
with asset returns have no effect on asset pricing, and shocks correlated
with asset returns are quickly traded away. 

The way around this problem is to make the idiosyncratic shocks permanent.
We can give individuals idiosyncratic income shocks that are correlated
with the market but are uncorrelated with returns. We can give people
income shocks that are uncorrelated with returns, so they cannot be
traded away. Then we exploit the non-linearity of marginal utility.
Then we have a nonlinear marginal utility function turn these shocks
into marginal utility shocks that are correlated with asset returns,
and hence can affect pricing implications. This is why Constantinides
and Duffie specify that the variance of idiosyncratic risk rises when
the market declines. If marginal utility were linear, an increase
in variance would have no effect on the average level of marginal
utility. Therefore, Constantinides and Duffie specify power utility,
and the interaction of nonlinear marginal utility and changing conditional
variance produces an equity premium. 
\end{doublespace}

Each consumer $i$ has power utility and a simple model can be specified
wherein, individual consumption growth $C_{it+1}$ is determined by
an independent idiosyncratic shock $\eta_{it}$ (Appendix \ref{subsec:Heterogeneous-Agents}
has the model specification and related details). The cross-sectional
standard deviation of consumption growth is specified so that people
suffer a high cross-sectional variance of consumption growth on dates
of a low market return. The excess return, $R_{t+1}^{e}$, can be
written, after aggregating across all consumers as,
\begin{equation}
0=E_{t}\left[\left(e^{-\alpha E_{N}\left[\Delta c_{it+1}\right]+\frac{\alpha^{2}}{2}\sigma_{N}^{2}\left[\Delta c_{it+1}\right]}\right)R_{t+1}^{e}\right]
\end{equation}
From this we see that the economy displays more risk aversion than
would a representative agent with aggregate consumption, $\triangle c_{t+1}^{a}=E_{N}\triangle c_{it+1}$.
If $\sigma_{N}$, the aggregate standard deviation of consumption
growth over all consumers $N$, varies over time, the risk aversion
can also vary over time and this variation can generate risk premia. 
\begin{doublespace}

\subsection{Unintended Consequences}
\end{doublespace}

\begin{doublespace}
Due to the dynamic nature of social systems, changes can be observed
and decisions effected by participants to influence the system. In
the social sciences, as soon as any generalization and its set of
conditions becomes common knowledge, the entry of many participants
shifts the equilibrium or the dynamics, such that the generalization
no longer applies to the known set of conditions. As long as participants
are free to observe the results and modify their actions, this effect
will persist and the varying behavior of participants in a social
system will give rise to unintended consequences. (Kashyap 2015; 2016)
discuss recent examples in the financial markets where unintended
consequences set in.

All attempts at prediction, including both the physical and the social
sciences, are like driving cars with the front windows blackened out.
The Uncertainty Principle of the Social Sciences can be stated as,
\textquotedblleft Any generalization in the social sciences cannot
be both popular and continue to yield accurate predictions or in other
words, the more popular a particular generalization, the less accurate
will be the predictions it yields\textquotedblright . An artifact
of this is unintended consequences. Many long standing puzzles seem
to have been resolved using different techniques. The various explanations
are still to be tested over time before acceptance; but then unexpected
outcomes set in and new puzzles emerge. As real analysis and limits
tell us (Rosenlicht 1968; Schumacher 2008): We are getting Closer
and Closer; Yet it seems we are still Far Far Away... 
\end{doublespace}

(McManus and Hastings 2005) clarify the wide range of uncertainties
that affect complex engineering systems and present a framework to
understand the risks (and opportunities) they create and the strategies
system designers can use to mitigate or take advantage of them. (Simon
1962) points out that any attempt to seek properties common to many
sorts of complex systems (physical, biological or social), would lead
to a theory of hierarchy since a large proportion of complex systems
observed in nature exhibit hierarchic structure. (Lawson 1985) argues
that the Keynesian view on uncertainty (that it is generally impossible,
even in probabilistic terms, to evaluate the future outcomes of all
possible current actions; Keynes 1937; 1971; 1973), far from being
innocuous or destructive of economic analysis in general, can give
rise to research programs incorporating, amongst other things, a view
of rational behavior under uncertainty, which could be potentially
fruitful. These viewpoints hold many lessons for policy designers
in the social sciences and could be instructive for researchers looking
at ways to understand and contend with complex systems, keeping in
mind the caveats of dynamic social systems.

\begin{doublespace}
Another under-appreciated problem in empirical results on asset pricing
is that a large part of the U.S. postwar average stock return may
represent good luck rather than ex ante expected return. The standard
deviation of stock returns is so high that standard errors are surprisingly
large that an eight percent return is not statistically different
from zero. (Siegel 1992a, b) extend the U.S. data on real stock and
bond returns back to 1802 and find that early stock returns did not
exceed fixed income returns by nearly the same magnitude they did
in more recent data. The equity premium is a puzzle because the measured
risk associated with equity returns is not high enough to justify
the observed high returns. (Poterba and Summers 1988) show that the
standard deviation of stock returns actually decreases more quickly
than it would if returns were a random walk because stock returns
display mean reversion. (Siegel and Thaler 1997) highlight that asset
returns deviate from a random walk, which implies that for long-horizon
investors, the risk of holding stocks is less than one would expect
by just looking at the annual standard deviation of returns. 
\end{doublespace}
\begin{doublespace}

\subsection{Transversality Condition}
\end{doublespace}

\begin{doublespace}
\begin{equation}
\underset{t\rightarrow\infty}{\lim}E_{t}\left[m_{t,t+j}p_{t+j}\right]=0
\end{equation}

This innocuous assumption is made to rule out the formation of asset
pricing bubbles, so that prices grow so fast that people will buy
now just to resell at higher prices later, even if there are no dividends.
The reality of financial markets makes it clear that participants
trade primarily to benefit from temporary bubbles or to capitalize
from a jump in prices. Hence, we need to consider if there are any
alternatives to this assumption or can this be relaxed under any situations?

We need to consider the extent of trading in stocks as compared to
other assets, or the risk free asset. The equity premium could be
due to the possibility that stocks are available for trading by a
larger segment of the population and there is a possibility that the
equity market can harbor price bubbles more than any other asset class.
If the expectations of investors change in such a way that they believe
they will be able to sell an asset for a higher price in the future
than they had been expecting, then the current price of the asset
will rise (Stiglitz 1990). If the reason that the price is high today
is only because investors believe that the selling price will be high
tomorrow-when \textquotedbl fundamental\textquotedbl{} factors do
not seem to justify such a price-then a bubble exists. If the asset
price increases more slowly than the discount factor, eventually the
terminal price becomes of negligible importance as viewed from today.
Under such circumstances, the value of the asset has to be just equal
to the discounted value of the stream of returns it generates, and
no bubbles can exist. But as long as no one in the economy has an
infinite planning horizon, there is nothing to ensure that this condition
on prices (called the transversality condition) will be satisfied.

(Weitzman 1973; Araujo and Scheinkman 1983) derive duality conditions
necessary and sufficient for infinite horizon optimality, emphasizing
the close connection between duality theory for infinite horizon convex
models and dynamic programming, showing that a necessary and sufficient
condition for optimality is the existence of support prices such that
the limit value of the optimal capital stocks is zero. (Michel 1982;
Benveniste and Scheinkman 1982) discuss the assumptions required to
set to zero the value of the stocks at the limit. (Michel 1990) studies
general concave discrete time infinite horizon optimal control problem
and establish necessary and sufficient conditions for optimality.
In finite horizon optimal control problems without constraints on
the final state, necessary conditions for optimality include the transversality
condition: the final value of the shadow price-vector is zero. This
means that one more unit of any good at final time gives no additional
value to the criterion. Halkin's example (Halkin 1974) shows that
this property is not necessarily true in an infinite horizon. In an
infinite horizon, one more unit of a good, at any time, changes the
whole future, and the zero value of the state becomes a limit property
which is not necessarily verified. (Ekeland and Scheinkman 1986) prove
the necessity of a standard transversality condition under certain
technical conditions; (Kamihigashi 2000) provides a simplification
of the same proof with some relaxed assumptions. (Benveniste and Scheinkman
1982) prove the envelope condition to find the derivative of the value
function of a recursive optimization problem. (Kamihigashi 2000; 2001;
2002) provide proofs for the necessity of transversality conditions
under deterministic scenarios. (Kamihigashi 2003) considers stochastic
versions.
\end{doublespace}
\begin{doublespace}

\subsection{How Durable is Durable? Nothing Lasts Forever... }
\end{doublespace}

\begin{doublespace}
The equity premium puzzle is based on the consumption growth of non-durable
goods and services. (Startz 1989) looks at the time series behavior
of consumption and verifies that purchases of non-durable goods follow
a random walk while purchases of durable goods require an ARMAX model
(X-extension of autoregressive\textendash moving-average, ARMA, models
with X-exogenous inputs; See Hamilton 1994), one in which lags of
nondurables and services enter on the right-hand side. (Conrad and
Schr�der 1991) propose an integrated framework for modeling consumer
demand for durables and nondurables and employ this approach for measuring
the effect of an enforced environmental policy on energy demand and
on consumer welfare. (Erceg and Levin 2002) find that a monetary policy
innovation has a peak impact on durable expenditures that is several
times as large as its impact on non-durable expenditures, and hence
a greater interest rate sensitivity.

Looking at the companies listed on the NASDAQ (End-notes \ref{enu:Durable}
and \ref{enu:Non-Durable}), we see that a huge number of companies
are labeled as durable goods producers. The aggregate valuation of
durable goods providers is comparable to the aggregate valuation of
non-durable producers. As of May 24, 2016 of the companies listed
on the NASDAQ, 235 were non-durable producers with 2.5 Trillion USD
market capitalization versus 151 durable producers with 400 Billion
USD market capitalization; worth noting that the top ten non-durable
companies have a combined market cap of 1.4 Trillion USD. The profits
and cash flow from these two groups needs to be analyzed further.
The change in consumption or spending can be argued to be higher for
durable goods than for non-durable goods, since most basic necessities
fall under non-durable goods. Do housing prices change more during
bad times as compared to the price of toothpaste and milk? If we look
at consumption changes in either group separately, then perhaps we
need to consider returns from the stock market for each group separately
as well. 
\end{doublespace}
\begin{doublespace}

\section{Conclusions }
\end{doublespace}

\begin{doublespace}
We have discussed the equity premium puzzle and a few well known attempts
to resolve it. Additional state variables are the natural route to
solving empirical puzzles. The Campbell\textendash Cochrane model
is a representative from the literature that attacks the equity premium
by modifying the representative agent\textquoteright s preferences.
The Constantinides and Duffie model is a representative of the literature
that attacks the equity premium by modeling uninsured idiosyncratic
risks, market frictions, and limited participation. These models are
quite similar in spirit. First, both models make a similar, fundamental
change in the description of stock market risk. Consumers do not fear
much the loss of wealth of a bad market return per se. They fear that
loss of wealth because it tends to come in recessions, in one case
defined as times of heightened idiosyncratic labor market risk, and
in the other case defined as a fall of consumption relative to its
recent past. This recession state variable or risk factor drives most
variation in expected returns. The Banson and Yaron model modifies
the representative agent\textquoteright s preferences and separates
the inter-temporal elasticity of substitution and the risk aversion
parameter and introduces variables to capture long term uncertainty.
All these models are an artifact of having many parameters in the
model so that some parameters can be calibrated to explain particular
facets of a phenomenon and other parameters can be set to explain
related but different facets of the same phenomenon.

We have discussed some possibilities for future research that might
be able to resolve this and other related puzzles in the social sciences.
Many long standing puzzles seem to have been resolved using different
techniques. The various explanations need to stand the test of time
before acceptance; but then unexpected outcomes set in and new puzzles
emerge. As real analysis and limits tell us: we are getting Closer
and Closer; yet it seems we are still Far Far Away...
\end{doublespace}
\begin{doublespace}

\section{Acknowledgements and End-notes}
\end{doublespace}
\begin{enumerate}
\item The author is grateful to Dr. Srikant Marakani for his patient clarifications,
useful suggestions and many interesting discussions on finance and
other equally esoteric topics. Dr. Yong Wang, Dr. Isabel Yan, Dr.
Vikas Kakkar, Dr. Fred Kwan, Dr. William Case, Dr. Costel Daniel Andonie,
Dr. Guangwu Liu, Dr. Jeff Hong, Dr. Andrew Chan, Dr. Humphrey Tung
and Dr. Xu Han at the City University of Hong Kong provided advice
and more importantly encouragement to explore and where possible apply
cross disciplinary techniques. The views and opinions expressed in
this article, along with any mistakes, are mine alone and do not necessarily
reflect the official policy or position of either of my affiliations
or any other agency.
\item \label{enu:Unlimited Limited}Economics is a social science that studies
how people satisfy unlimited wants with scarce resources. It involves
the analysis of choice and trade through the use of intuitive graphs
and mathematical elements. \href{https://en.wikibooks.org/wiki/Principles_of_Economics/What_Is_Economics}{Unlimited Wants Limited Resources}
\begin{doublespace}
\item \label{enu:Durable}List of Consumer Durables Companies with Market
Capitalization, Stock Symbol, Country and Industrial Subsector on
NASDAQ: \href{http://www.nasdaq.com/screening/companies-by-industry.aspx?industry=Consumer+Durables}{Consumer Durables Companies on NASDAQ}
\item \label{enu:Non-Durable}List of Consumer Non-Durables Companies with
Market Capitalization, Stock Symbol, Country and Industrial Subsector
on NASDAQ: \href{http://www.nasdaq.com/screening/companies-by-industry.aspx?industry=Consumer+Non-Durables}{Consumer Non-Durables Companies on NASDAQ}
\end{doublespace}
\end{enumerate}

\section{References}
\begin{enumerate}
\begin{doublespace}
\item Altug, S. (1983). Gestation lags and the business cycle. Presented
at the 1984 summer meetings of the Econometric Society. Carnegie-Mellon
University.
\item Araujo, A., \& Scheinkman, J. A. (1983). Maximum principle and transversality
condition for concave infinite horizon economic models. Journal of
Economic Theory, 30(1), 1-16.
\item Arrow, K. J. (1971). Essays in the theory of risk-bearing (North-Holland,
Amsterdam).
\end{doublespace}
\item Arrow, K. J., \& Debreu, G. (1954). Existence of an Equilibrium for
a Competitive Economy. Econometrica, 22(3), 265-290.
\begin{doublespace}
\item Bansal, R., \& Yaron, A. (2004). Risks for the long run: A potential
resolution of asset pricing puzzles. The Journal of Finance, 59(4),
1481-1509.
\end{doublespace}
\item Bansal, R., Dittmar, R. F., \& Lundblad, C. T. (2005). Consumption,
dividends, and the cross section of equity returns. The Journal of
Finance, 60(4), 1639-1672.
\item Bansal, R., Kiku, D., \& Yaron, A. (2010). Long run risks, the macroeconomy,
and asset prices. The American Economic Review, 100(2), 542-546.
\begin{doublespace}
\item Barro, R. J. (2006). Rare disasters and asset markets in the twentieth
century. The Quarterly Journal of Economics, 823-866.
\item Barrow, J. D. (1991). Theories of everything: The quest for ultimate
explanation.
\item Barrow, J. D. (2007). New theories of everything. OUP Oxford.
\item Benveniste, L. M., \& Scheinkman, J. A. (1982). Duality theory for
dynamic optimization models of economics: The continuous time case.
Journal of Economic Theory, 27(1), 1-19.
\end{doublespace}
\item Brown, S. J., Goetzmann, W. N., \& Ross, S. A. (1995). Survival. The
Journal of Finance, 50(3), 853-873.
\begin{doublespace}
\item Campbell, J. Y., \& Cochrane, J. H. (1999). By Force of Habit: A Consumption-Based
Explanation of Aggregate Stock Market Behavior. The Journal of Political
Economy, 107(2), 205-251.
\end{doublespace}
\item Campbell, J. Y., \& Shiller, R. J. (1988). The Dividend-Price Ratio
and Expectations of Future Dividends and Discount Factors. The Review
of Financial Studies, 1(3), 195-228.
\begin{doublespace}
\item Cao, Z., Cao, H. G., \& Qiang, W. (2015). Theory of Everything. Frontiers
of Astronomy Astrophysics and Cosmology, 1(1), 31-36.
\item Cao, Z., \& Cao, H. G. (2013). Unified field theory. American Journal
of Modern Physics, 2(6), 292-298.
\end{doublespace}
\item Chen, X., Favilukis, J., \& Ludvigson, S. C. (2013). An estimation
of economic models with recursive preferences. Quantitative Economics,
4(1), 39-83.
\begin{doublespace}
\item Cochrane, J. H. (2009). Asset Pricing:(Revised Edition). Princeton
university press.
\end{doublespace}
\item Conrad, K., \& Schr�der, M. (1991). Demand for durable and non-durable
goods, environmental policy and consumer welfare. Journal of Applied
Econometrics, 6(3), 271-286.
\begin{doublespace}
\item Constantinides, G. M. (1982). Intertemporal asset pricing with heterogeneous
consumers and without demand aggregation. Journal of Business, 253-267.
\end{doublespace}
\item Constantinides, G. M. (1990). Habit Formation: A Resolution of the
Equity Premium Puzzle. The Journal of Political Economy, 98(3), 519-543.
\begin{doublespace}
\item Constantinides, G. M., \& Duffie, D. (1996). Asset pricing with heterogeneous
consumers. Journal of Political economy, 219-240.
\item Courant, R., \& Robbins, H. (1996). What is Mathematics?: an elementary
approach to ideas and methods. Oxford University Press.
\end{doublespace}
\item Debreu, G. (1987). Theory of value: An axiomatic analysis of economic
equilibrium (Vol. 17). Yale University Press.
\item Drechsler, I., \& Yaron, A. (2011). What's vol got to do with it.
Review of Financial Studies, 24(1), 1-45.
\item Dusenberry, J. S. (1949). Income, Savings and the Theory of Consumer
Behavior. Harvard University Press, Cambridge, MA.
\begin{doublespace}
\item Ekeland, I., \& Scheinkman, J. A. (1986). Transversality conditions
for some infinite horizon discrete time optimization problems. Mathematics
of operations research, 11(2), 216-229.
\item Epstein, L. G., \& Zin, S. E. (1989). Substitution, risk aversion,
and the temporal behavior of consumption and asset returns: A theoretical
framework. Econometrica: Journal of the Econometric Society, 937-969.
\end{doublespace}
\item Erceg, C. J., \& Levin, A. T. (2002). Optimal monetary policy with
durable and non-durable goods. FRB International Finance Discussion
Paper, (748).
\item Fama, E. F., \& French, K. R. (1993). Common risk factors in the returns
on stocks and bonds. Journal of financial economics, 33(1), 3-56.
\item Fama, E. F., \& French, K. R. (2002). The equity premium. The Journal
of Finance, 57(2), 637-659.
\begin{doublespace}
\item Friend, I., \& Blume, M. E. (1975). The demand for risky assets. The
American Economic Review, 65(5), 900-922.
\item Halkin, H. (1974). Necessary Conditions for Optimal Control Problems
with Infinite Horizons. Econometrica, 42(2), 267-72.
\end{doublespace}
\item Hamilton, J. D. (1994). Time series analysis (Vol. 2). Princeton:
Princeton university press.
\begin{doublespace}
\item Hansen, L. P., \& Jagannathan, R. (1991). Implications of Security
Market Data for Models of Dynamic Economies. The Journal of Political
Economy, 99(2), 225-262.
\item Hildreth, C., \& Knowles, G. J. (1982). Some estimates of Farmers'
utility functions (No. 54545). University of Minnesota, Agricultural
Experiment Station.
\item Hull, J. C. (2010). Options, Futures, and Other Derivatives, 7/e (With
CD). Pearson Education India.
\end{doublespace}
\item Jagannathan, R., \& Marakani, S. (2015). Price-Dividend Ratio Factor
Proxies for Long-Run Risks. Review of Asset Pricing Studies, 5(1),
1-47.
\begin{doublespace}
\item Kamihigashi, T. (2000). A simple proof of Ekeland and Scheinkman's
result on the necessity of a transversality condition. Economic Theory,
15(2), 463-468.
\item Kamihigashi, T. (2001). Necessity of transversality conditions for
infinite horizon problems. Econometrica, 69(4), 995-1012.
\item Kamihigashi, T. (2002). A simple proof of the necessity of the transversality
condition. Economic theory, 20(2), 427-433.
\item Kamihigashi, T. (2003). Necessity of transversality conditions for
stochastic problems. Journal of Economic Theory, 109(1), 140-149.
\item Kashyap, R. (2014). Dynamic Multi-Factor Bid\textendash Offer Adjustment
Model. The Journal of Trading, 9(3), 42-55.
\item Kashyap, R. (2015). A Tale of Two Consequences. The Journal of Trading,
10(4), 51-95.
\end{doublespace}
\item Kashyap, R. (2016). Hong Kong - Shanghai Connect / Hong Kong - Beijing
Disconnect (?), Scaling the Great Wall of Chinese Securities Trading
Costs. The Journal of Trading, 11(3), 81-134.
\begin{doublespace}
\item Kehoe, P. J., \& Richardson, P. (1984). Dynamics of the current account:
Theoretical and empirical analysis. working paper, Harvard university.
\end{doublespace}
\item Keynes, J. M. (1937). The General Theory of Employment. The Quarterly
Journal of Economics, 51(2), 209-223.
\item Keynes, J. M. (1971). The Collected Writings of John Maynard Keynes:
In 2 Volumes. A Treatise on Money. The Applied Theory of Money. Macmillan
for the Royal Economic Society.
\item Keynes, J. M. (1973). A treatise on probability, the collected writings
of John Maynard Keynes, vol. VIII.
\begin{doublespace}
\item Kydland, F., \& Prescott, E. (1982). Time to Build and Aggregate Fluctuations.
Econometrica, 50(6), 1345-70.
\end{doublespace}
\item Lawson, T. (1985). Uncertainty and economic analysis. The Economic
Journal, 95(380), 909-927.
\begin{doublespace}
\item Lucas Jr, R. E. (1978). Asset prices in an exchange economy. Econometrica:
Journal of the Econometric Society, 1429-1445.
\item Laughlin, R. B., \& Pines, D. (2000). The theory of everything. Proceedings
of the National Academy of Sciences, 97(1), 28-31.
\end{doublespace}
\item Marshall, A. Principles of Economics: An Introductory Volume. 8th
edition. London: Macmillan, 1920.
\item McKenzie, L. W. (1954). On equilibrium in Graham's model of world
trade and other competitive systems. Econometrica: Journal of the
Econometric Society, 22 (2), 147-161.
\item McKenzie, L. W. (1959). On the existence of general equilibrium for
a competitive market. Econometrica: journal of the Econometric Society,
27 (1), 54-71.
\item McManus, H., \& Hastings, D. (2005, July). 3.4. 1 A Framework for
Understanding Uncertainty and its Mitigation and Exploitation in Complex
Systems. In INCOSE International Symposium (Vol. 15, No. 1, pp. 484-503).
\item Merton, R. C. (1973). An Intertemporal Capital Asset Pricing Model.
Econometrica: Journal of the Econometric Society, 41(5), 867-87.
\begin{doublespace}
\item Mehra, R., \& Prescott, E. C. (1985). The equity premium: A puzzle.
Journal of monetary Economics, 15(2), 145-161.
\item Mehra, R. (1988). On the existence and representation of equilibrium
in an economy with growth and non-stationary consumption. International
Economic Review, 131-135.
\item Mehra, R., \& Prescott, E. C. (1988). The equity risk premium: A solution?.
Journal of Monetary Economics, 22(1), 133-136.
\item Michel, P. (1982). On the Transversality Condition in Infinite Horizon
Optimal Problems. Econometrica, 50(4), 975-985.
\item Michel, P. (1990). Some Clarifications on the Transversality Condition.
Econometrica, 58(3), 705-23.
\end{doublespace}
\item Pollak, R. A. (1970). Habit formation and dynamic demand functions.
Journal of political Economy, 78(4), 745-763.
\item Poterba, J. M., \& Summers, L. H. (1988). Mean reversion in stock
prices: Evidence and implications. Journal of financial economics,
22(1), 27-59.
\begin{doublespace}
\item Rietz, T. A. (1988). The equity risk premium a solution. Journal of
monetary Economics, 22(1), 117-131.
\item Rosenlicht, M. (1968). Introduction to analysis. Courier Corporation.
\item Royden, H. L., \& Fitzpatrick, P. (1988). Real analysis (Vol. 198,
No. 8). New York: Macmillan.
\item Rudin, W. (1964). Principles of mathematical analysis (Vol. 3). New
York: McGraw-Hill.
\item Schumacher, C. (2008). Closer and closer: Introducing real analysis.
Jones and Bartlett Publishers.
\end{doublespace}
\item Siegel, J. J. (1992a). The equity premium: Stock and bond returns
since 1802. Financial Analysts Journal, 48(1), 28-38.
\item Siegel, J. J. (1992b). The real rate of interest from 1800\textendash 1990:
A study of the US and the UK. Journal of Monetary Economics, 29(2),
227-252.
\item Siegel, J. J., \& Thaler, R. H. (1997). Anomalies: The equity premium
puzzle. The Journal of Economic Perspectives, 11(1), 191-200.
\item Simon, H. A. (1962). The Architecture of Complexity. Proceedings of
the American Philosophical Society, 106(6), 467-482.
\item Startz, R. (1989). The Stochastic Behavior of Durable and Non-durable
Consumption. The Review of Economics and Statistics, 71(2), 356-63.
\begin{doublespace}
\item Stiglitz, J. E. (1990). Symposium on bubbles. The Journal of Economic
Perspectives, 4(2), 13-18.
\item Taleb, N. N. (2007). The black swan: The impact of the highly improbable.
Random House.
\item Tegmark, M. (1998). Is \textquotedblleft the theory of everything\textquotedblright{}
merely the ultimate ensemble theory?. Annals of Physics, 270(1), 1-51.
\item Tegmark, M. (2008). The mathematical universe. Foundations of Physics,
38(2), 101-150.
\item Tobin, J., \& Dolde, W. (1971). Wealth, liquidity and consumption.
Consumer spending and monetary policy: The linkage (Federal Reserve
Bank of Boston, Boston, MA), 5, 99-160.
\item Weil, P. (1989). The equity premium puzzle and the risk-free rate
puzzle. Journal of Monetary Economics, 24(3), 401-421.
\item Weil, P. (1990). Non-expected Utility in Macroeconomics. The Quarterly
Journal of Economics, 105(1), 29-42.
\item Weitzman, M. L. (1973). Duality theory for infinite horizon convex
models. Management Science, 19(7), 783-789.
\item Weitzman, M. (2007). Subjective Expectations and Asset-Return Puzzles.
American Economic Review, 97(4), 1102-1130.
\end{doublespace}
\end{enumerate}

\section{Appendix: Mathematical Steps}
\begin{doublespace}

\subsection{\label{subsec:Hansen-and-Jagannathan}Hansen and Jagannathan Bound}
\end{doublespace}

\begin{doublespace}
All assets priced by the discount factor $m$ need to be obey (Cochrane
2009),
\[
1=E\left(mR^{i}\right)
\]
\[
1=E\left(m\right)E\left(R^{i}\right)+\rho_{m,R^{i}}\sigma\left(R^{i}\right)\sigma\left(m\right)
\]
\[
E\left(R^{i}\right)=R^{f}-\rho_{m,R^{i}}\frac{\sigma\left(m\right)}{E\left(m\right)}\sigma\left(R^{i}\right)\quad\because\frac{1}{E\left(m\right)}=R^{f}
\]
\[
\left|\frac{E\left(R^{i}\right)-R^{f}}{\sigma\left(R^{i}\right)}\right|\leq\frac{\sigma\left(m\right)}{E\left(m\right)}\quad\because\left|\rho_{m,R^{i}}\right|\leq1
\]
We could also write this as,
\[
1=E\left(mR^{i}\right)\Rightarrow e^{0}=E\left[e^{\ln m+\ln R^{i}}\right]=e^{E\left(\ln m\right)+E\left(\ln R^{i}\right)+\frac{1}{2}\text{Var}\left(\ln m+\ln R^{i}\right)}
\]
\[
0=E\left(\ln m\right)+E\left(\ln R^{i}\right)+\frac{1}{2}\text{Var}\left(\ln m\right)+\frac{1}{2}\text{Var}\left(\ln R^{i}\right)+\text{Cov}\left(\ln m,\ln R^{i}\right)\quad\because\text{Var}\left(X+Y\right)=\text{Var}\left(X\right)+\text{Var}\left(Y\right)+2\text{Cov}\left(X,Y\right)
\]
\[
0=E\left(\ln R^{i}\right)+\frac{1}{2}\text{Var}\left(\ln R^{i}\right)-\ln R^{f}+\text{Cov}\left(\ln m,\ln R^{i}\right)
\]
\begin{eqnarray*}
\left[\because Y=\ln X\sim N\left\{ \mu=E\left(Y\right)=E\left(\ln X\right),\sigma^{2}=\text{Var}\left(Y\right)=\text{Var}\left(\ln X\right)\right\} \Rightarrow E\left(X\right)=e^{\mu+\frac{1}{2}\sigma^{2}}\Rightarrow\ln E\left(X\right)=\mu+\frac{1}{2}\sigma^{2}\right.\\
\left.\text{Above, set }X=m\Rightarrow E\left(\ln m\right)+\frac{1}{2}\text{Var}\left(\ln m\right)=\ln E\left(m\right);\quad\ln E\left(m\right)=\ln\frac{1}{R^{f}}=-\ln R^{f}\right]
\end{eqnarray*}
For an excess return these equations become, 
\[
0=E\left(mR^{e}\right)
\]
\[
0=E\left(m\right)E\left(R^{e}\right)+\rho_{m,R^{e}}\sigma\left(R^{e}\right)\sigma\left(m\right)
\]
\[
\left|\frac{E\left(R^{e}\right)}{\sigma\left(R^{e}\right)}\right|\leq\frac{\sigma\left(m\right)}{E\left(m\right)}
\]
When the correlation $\left|\rho_{m,R^{mv}}\right|=1$ or for assets
on the mean variance frontier we have,
\[
\left|\frac{E\left(R^{mv}\right)-R^{f}}{\sigma\left(R^{mv}\right)}\right|=\frac{\sigma\left(m\right)}{E\left(m\right)}
\]
Assuming a power utility function of the constant relative risk aversion
class of the form, 
\[
U\left(C_{t},\alpha\right)=\frac{C_{t}^{1-\alpha}-1}{1-\alpha}
\]
Here, $\alpha$ measures the curvature of the utility function and
$C_{t}$ is the consumption at time $t$. This specification ensures
that the equilibrium return process is stationary. We have from the
first order conditions of a utility maximizing representative consumer,
\[
m=\beta\frac{U'\left(C_{t+1}\right)}{U'\left(C_{t}\right)}
\]
\[
\left|\frac{E\left(R^{mv}\right)-R^{f}}{\sigma\left(R^{mv}\right)}\right|=\frac{\sigma\left[\left(\frac{C_{t+1}}{C_{t}}\right)^{-\alpha}\right]}{E\left[\left(\frac{C_{t+1}}{C_{t}}\right)^{-\alpha}\right]}
\]
Assuming that consumption growth is log normal, i.e., $\Delta c_{t+1}=\ln C_{t+1}-\ln C_{t}=\ln\left(\frac{C_{t+1}}{C_{t}}\right)=x$,
say, which is a normally distributed random variable. 
\[
E\left(e^{x}\right)=e^{E\left(x\right)+\frac{1}{2}\sigma^{2}\left(x\right)}
\]
\[
\sigma^{2}\left[\left(\frac{C_{t+1}}{C_{t}}\right)^{-\alpha}\right]=E\left\{ \left[\left(\frac{C_{t+1}}{C_{t}}\right)^{-\alpha}\right]^{2}\right\} -\left\{ E\left[\left(\frac{C_{t+1}}{C_{t}}\right)^{-\alpha}\right]\right\} ^{2}
\]
\[
\sigma\left[\left(\frac{C_{t+1}}{C_{t}}\right)^{-\alpha}\right]=\sqrt{E\left[e^{-2\alpha x}\right]-\left\{ E\left[e^{-\alpha x}\right]\right\} ^{2}}
\]
For any real or complex number $s$, the $s$-th moment of a log-normally
distributed variable $X\equiv e^{Y}$ is given by,
\[
\operatorname{E}[X^{s}]=e^{sE\left(Y\right)+\frac{1}{2}s^{2}\sigma^{2}\left(Y\right)}
\]
\[
\sigma\left[\left(\frac{C_{t+1}}{C_{t}}\right)^{-\alpha}\right]=\sqrt{e^{-2\alpha E\left(x\right)+2\alpha^{2}\sigma^{2}\left(x\right)}-\left\{ e^{-\alpha E\left(x\right)+\frac{1}{2}\alpha^{2}\sigma^{2}\left(x\right)}\right\} ^{2}}
\]
\[
=\sqrt{e^{-2\alpha E\left(x\right)+\alpha^{2}\sigma^{2}\left(x\right)}\left[e^{\alpha^{2}\sigma^{2}\left(x\right)}-1\right]}
\]
\[
=e^{-\alpha E\left(x\right)+\frac{1}{2}\alpha^{2}\sigma^{2}\left(x\right)}\sqrt{\left[e^{\alpha^{2}\sigma^{2}\left(x\right)}-1\right]}
\]
\[
\sigma\left[\left(\frac{C_{t+1}}{C_{t}}\right)^{-\alpha}\right]=E\left[\left(\frac{C_{t+1}}{C_{t}}\right)^{-\alpha}\right]\sqrt{\left[e^{\alpha^{2}\sigma^{2}\left(x\right)}-1\right]}
\]
\[
\left|\frac{E\left(R^{mv}\right)-R^{f}}{\sigma\left(R^{mv}\right)}\right|=\sqrt{\left[e^{\alpha^{2}\sigma^{2}\left(x\right)}-1\right]}
\]
If $\alpha^{2}\sigma^{2}\left(x\right)$ is small, then using the
approximation, $e^{y}\approx1+y$, the following holds,
\[
\left|\frac{E\left(R^{mv}\right)-R^{f}}{\sigma\left(R^{mv}\right)}\right|\approx\alpha\sigma\left(\Delta c_{t+1}\right)
\]

\end{doublespace}

\subsection{\label{subsec:Consumption-Growth-and}Consumption Growth and Interest
Rates}

\begin{doublespace}
\[
R^{f}=\frac{1}{E\left(m\right)}=\frac{1}{E\left[\beta\left(\frac{c_{t+1}}{c_{t}}\right)^{-\alpha}\right]}
\]
Similar to before, if we have log normal consumption growth, $\Delta c_{t+1}=\ln\left(\frac{C_{t+1}}{C_{t}}\right)=x$
and setting $\beta=e^{-\delta}$, 
\[
R^{f}=\left[e^{-\delta}e^{-\alpha E\left(x\right)+\frac{1}{2}\alpha^{2}\sigma^{2}\left(x\right)}\right]^{-1}
\]
Taking logarithms,
\[
r^{f}=\ln R^{f}=\delta+\alpha E\left(\Delta c_{t+1}\right)-\frac{\alpha^{2}}{2}\sigma^{2}\left(\Delta c_{t+1}\right)
\]

\end{doublespace}

\subsection{\label{subsec:Mehra-and-Prescott}Mehra and Prescott variation with
Markov Process}

The firm's output is constrained to be less than or equal to $y_{t}$.
$y_{t}$ is the firm's dividend payment in the period $t$ as well.
The growth rate in $y_{t}$ is subject to a Markov chain, 
\[
y_{t+1}=x_{t+1}y_{t}
\]
Here, $x_{t+1}\in\left\{ \lambda_{1},\ldots,\lambda_{n}\right\} $
is the growth rate and the transition probability between states $i$
and $j$, $\Pr\left\{ x_{t+1}=\lambda_{j};x_{t}=\lambda_{i}\right\} =\phi_{ij}$.
The price of any security in period $t$ with payments given by the
process $d_{s}$ is,
\[
P_{t}=E_{t}\left\{ \sum_{s=t+1}^{\infty}\beta^{s-t}\frac{U'\left(y_{s}\right)d_{s}}{U'\left(y_{t}\right)}\right\} 
\]
For an equity share of the firm with dividend payment process, $\left\{ y_{s}\right\} $
\[
P_{t}^{e}=P^{e}\left(x_{t},y_{t}\right)
\]
\[
=E_{t}\left\{ \sum_{s=t+1}^{\infty}\beta^{s-t}\frac{y_{t}^{\alpha}y_{s}}{y_{s}^{\alpha}}\left|x_{t,}y_{t}\right.\right\} 
\]
Under equilibrium, since, $y_{s}=y_{t}x_{t+1}\ldots x_{s}$, the price
is homogeneous of degree one in $y_{t}$, which is the current endowment
of the consumption good. The state is fully represented by $\left(x_{t},y_{t}\right)$.
Recognizing that the equilibrium values are time invariant functions
of the state, we can then redefine it as the pair $\left(c,i\right)\equiv\left(y_{t},\lambda_{i}\right)$.
The price of the equity share then satisfies (adopting small letters
for the prices and dropping time subscripts due to the time in-variance
of the functions under equilibrium), 
\[
p^{e}\left(c,i\right)=\beta\sum_{j=1}^{n}\phi_{ij}\left(\lambda_{j}c\right)^{-\alpha}\left[p^{e}\left(\lambda_{j}c,j\right)+\lambda_{j}c\right]c^{\alpha}
\]
Since, $p^{e}\left(c,i\right)$ is homogeneous of degree one in $c$,
we represent this function using a constant $w_{i}$ as, 
\[
p^{e}\left(c,i\right)=w_{i}c
\]
\[
\Rightarrow\qquad w_{i}c=\beta\sum_{j=1}^{n}\phi_{ij}\left(\lambda_{j}c\right)^{-\alpha}\left[\lambda_{j}w_{j}c+\lambda_{j}c\right]c^{\alpha}
\]
\[
\Rightarrow\qquad w_{i}=\beta\sum_{j=1}^{n}\phi_{ij}\left(\lambda_{j}\right)^{1-\alpha}\left[w_{j}+1\right]
\]
We then get the period return of the equity security as, 
\[
r_{ij}^{e}=\frac{p^{e}\left(\lambda_{j}c,j\right)+\lambda_{j}c-p^{e}\left(c,i\right)}{p^{e}\left(c,i\right)}=\frac{\lambda_{j}\left(w_{j}+1\right)}{w_{i}}-1
\]
The expected return denoted by capital letters with the current state
$i$, is,
\[
R_{i}^{e}=\sum_{j=1}^{n}\phi_{ij}r_{ij}^{e}
\]
Similarly we have for the risk free rate,
\[
p_{i}^{f}=p^{f}\left(c,i\right)=\beta\sum_{j=1}^{n}\phi_{ij}\left(\lambda_{j}\right)^{-\alpha}
\]
\[
R_{i}^{f}=\frac{1}{p_{i}^{f}}-1
\]
From the assumption of an ergodic Markov process, the vector of stationary
probabilities, $\pi\in R^{n}$ on state $i$ is given by the solution
of the system of equations, 
\[
\pi=\phi^{T}\pi\quad;\sum_{i=1}^{n}\pi_{i}=1\quad;\phi^{T}=\left\{ \phi_{ji}\right\} 
\]
The state independent returns for the equity and risk free security
and hence the equity risk premium are given by, 
\[
R^{e}=\sum_{i=1}^{n}\pi_{i}R_{i}^{e}\quad;\quad R^{f}=\sum_{i=1}^{n}\pi_{i}R_{i}^{f}
\]
With two states, the Markov process growth rates and transition probabilities
are,
\[
\lambda_{1}=1+\mu+\gamma,\quad\lambda_{2}=1+\mu-\gamma
\]
\[
\phi_{11}=\phi_{22}=\phi,\quad\phi_{12}=\phi_{21}=1-\phi
\]
The parameters $\left(\alpha,\beta\right)$ define preferences and
$\left(\mu,\phi,\gamma\right)$ define technology. They are estimated
using method of moments by matching the mean, variance and first order
auto-correlation of the growth rate of per-capita consumption. Based
on the estimated parameters, the maximum value of the equity premium
is 0.35 percent.

\subsection{\label{subsec:Force-of-Habit}Force of Habit}

\begin{doublespace}
Maximization of utility function now becomes,
\[
E\sum_{t=0}^{\infty}\beta^{t}\frac{\left(C_{t}-X_{t}\right)^{1-\alpha}-1}{1-\alpha}
\]
\[
\ln C_{t+1}-\ln C_{t}\equiv\Delta c_{t+1}=g+\vartheta_{t+1},\quad\vartheta_{t+1}\:\sim\:\text{i.i.d.}\;N\left(0,\sigma^{2}\right)
\]
Instead of the habit level, the log surplus consumption ratio, $s_{t}$,
evolves as a heteroskedastic AR(1) process,
\[
\ln S_{t+1}\equiv s_{t+1}=\left(1-\varrho\right)\bar{s}+\varrho s_{t}+\lambda\left(s_{t}\right)\left(c_{t+1}-c_{t}-g\right)
\]
Here, the surplus consumption ratio is given by,
\[
S_{t}=\frac{C_{t}-X_{t}}{C_{t}}
\]
The marginal utility is given by, 
\[
U_{c}\left(C_{t},X_{t}\right)=\left(C_{t}-X_{t}\right)^{-\alpha}=S_{t}^{-\alpha}C_{t}^{-\alpha}
\]
The inter-temporal marginal rate of substitution and hence the discount
factor are given by,
\[
M_{t+1}\equiv\beta\frac{U_{c}\left(C_{t+1},X_{t+1}\right)}{U_{c}\left(C_{t},X_{t}\right)}=\beta\left(\frac{S_{t+1}}{S_{t}}\frac{C_{t+1}}{C_{t}}\right)^{-\alpha}
\]
\[
M_{t+1}=\beta\left[e^{\left(1-\varrho\right)\left(\bar{s}-s_{t}\right)+\lambda\left(s_{t}\right)\left(c_{t+1}-c_{t}-g\right)}e^{g+\vartheta_{t+1}}\right]^{-\alpha}
\]
\[
=\beta G^{-\alpha}\left[e^{-\alpha\left\{ \left(1-\varrho\right)\left(\bar{s}-s_{t}\right)+\vartheta_{t+1}\left[1+\lambda\left(s_{t}\right)\right]\right\} }\right]
\]
\[
r_{t}^{f}=-\ln\left[E_{t}\left(M_{t+1}\right)\right]=-\ln\left(\beta\right)+\alpha g-\alpha\left(1-\varrho\right)\left(s_{t}-\bar{s}\right)-\frac{\alpha^{2}\sigma^{2}}{2}\left[1+\lambda\left(s_{t}\right)\right]^{2}
\]
The sensitivity function, $\lambda\left(s_{t}\right)$ is specified
to satisfy three conditions, 1) Risk free rate is constant; 2) habit
is predetermined at the steady state $\left(s_{t}-\bar{s}\right)$;
and 3) habit moves negatively with consumption everywhere, or equivalently,
habit is predetermined near the steady state. Simplifying gives, 
\[
\lambda\left(s_{t}\right)=\frac{1}{\bar{S}}\sqrt{1-2\left(s_{t}-\bar{s}\right)}-1\;;\quad\bar{S}=\sigma\sqrt{\frac{\alpha}{1-\varrho}}
\]
\[
r_{t}^{f}=-\ln\left(\beta\right)+\alpha g-\frac{\alpha}{2}\left(1-\varrho\right)
\]
Local curvature, $\eta_{t}$, depends on how far consumption is above
the habit, as well as $\alpha$,
\[
\eta_{t}=-\frac{C_{t}U_{cc}\left(C_{t}-X_{t}\right)}{U_{c}\left(C_{t}-X_{t}\right)}=\frac{\alpha}{S_{t}}
\]

\end{doublespace}
\begin{doublespace}

\subsection{\label{subsec:Long-Run-Risks}Long Run Risks and Survivors}
\end{doublespace}

\begin{doublespace}
The asset pricing restriction for gross return, $R_{i,t+1}$, satisfies,
\[
E_{t}\left[\beta^{\theta}G_{t+1}^{-\frac{\theta}{\varphi}}R_{a,t+1}^{-\left(1-\theta\right)}R_{i,t+1}\right]=1,\quad\theta=\frac{1-\alpha}{1-\frac{1}{\varphi}}
\]
Recursive preferences are given by, 
\[
U_{t}=\left\{ \left(1-\beta\right)C_{t}^{\frac{1-\alpha}{\theta}}+\beta E_{t}\left[U_{t+1}^{1-\alpha}\right]^{\frac{1}{\theta}}\right\} ^{\frac{\theta}{1-\alpha}}
\]
The logarithm of the inter-temporal marginal rate of substitution
is,
\[
\ln M_{t+1}\equiv m_{t+1}=\theta\ln\beta-\frac{\theta}{\varphi}g_{t+1}+\left(\theta-1\right)r_{a,t+1}
\]
Asset returns satisfy,
\[
E_{t}\left[e^{m_{t,t+1}+r_{i,t+1}}\right]=1
\]
$\varphi\geq0$ is the IES parameter. $G_{t+1}$ is the aggregate
growth rate of consumption; $R_{a,t+1}$ is the unobservable gross
return on an asset that delivers aggregate consumption as its dividend
each period; $R_{m,t+1}$ is the observable return on the market portfolio
and the return on the aggregate dividend claim; $g_{t+1}\;,r_{a,t+1}\;,\;r_{m,t+1}$
are the logarithms of the variables just mentioned. $q_{t}$ is a
small persistent predictable component in the consumption and dividend
growth rates. The growth rates on consumption $g_{t+1}$ and dividends
$g_{d,t+1}$ are modeled as below, (following Campbell and Shiller
1988, who use similar log linear approximations to show that price-dividend
ratios seem to predict long-horizon equity returns),
\[
\ln\left(R_{a,t+1}\right)=r_{a,t+1}=a_{0}+a_{1}z_{t+1}-z_{t}+g_{t+1},\quad z_{t}=\ln\left(\frac{P_{t}}{C_{t}}\right)
\]
\begin{eqnarray*}
q_{t+1} & = & \rho q_{t}+\varphi_{e}\sigma_{t}\varepsilon_{t+1}\\
g_{t+1} & = & \mu+q_{t}+\sigma_{t}\eta_{t+1}\\
g_{d,t+1} & = & \mu_{d}+\varphi q_{t}+\varphi_{d}\sigma_{t}u_{t+1}\\
\sigma_{t+1}^{2} & = & \sigma^{2}+\nu_{1}\left(\sigma_{t}^{2}-\sigma^{2}\right)+\sigma_{w}w_{t+1}\\
 &  & \varepsilon_{t+1},\eta_{t+1},u_{t+1},w_{t+1}\sim\text{i.i.d. }N\left(0,1\right)
\end{eqnarray*}
Here, $\sigma_{t+1}$, represents the time-varying economic uncertainty
incorporated in consumption growth rate and $\sigma^{2}$ is its unconditional
mean. There is an assumption that the shocks are uncorrelated, and
allow for only one source of economic uncertainty to affect consumption
and dividends. The risk free rate, $r_{f,t}$, and the equity premium,
$E_{t}\left(r_{m,t+1}-r_{f,t}\right)$, in the presence of time-varying
economic uncertainty are,
\[
r_{f,t}=-\theta\ln\beta+\frac{\theta}{\varphi}E_{t}\left[g_{t+1}\right]+\left(1-\theta\right)E_{t}\left[r_{a,t+1}\right]-\frac{1}{2}\text{var}_{t}\left[\frac{\theta}{\varphi}g_{t+1}+\left(1-\theta\right)r_{a,t+1}\right]
\]
\[
E_{t}\left(r_{m,t+1}-r_{f,t}\right)=\beta_{m,e}\lambda_{m,e}\sigma_{t}^{2}+\beta_{m,w}\lambda_{m,w}\sigma_{w}^{2}-0.5\text{var}_{t}\left(r_{m,t+1}\right)
\]
$\beta_{m,e}\:,\:\lambda_{m,e}\:,\:\beta_{m,w}\:,\:\lambda_{m,w}$
are combinations of other parameters. A simpler specification can
set $g_{t+1}=\mu+q_{t}+\sigma\eta_{t+1}$. But since the economic
uncertainty, $\sigma$, is constant, the conditional risk premium
and the conditional volatility of the market portfolio is constant
and hence their ratio, the Sharpe ratio is also constant. The long
run risk or time varying uncertainty gives a large value for the equity
premium while the separation between the IES parameter and risk aversion
ensures that the risk free rate remains small.
\end{doublespace}
\begin{doublespace}

\subsection{\label{subsec:Heterogeneous-Agents}Heterogeneous Agents}
\end{doublespace}

\begin{doublespace}
Each consumer $i$ has power utility, 

\[
U=E\sum_{t}e^{-\delta t}C_{it}^{1-\alpha}
\]
The simple model can be specified such that, individual consumption
growth $C_{it+1}$ is determined by an independent idiosyncratic shock
$\eta_{it}$, 
\[
\ln\left(\frac{C_{it+1}}{C_{it}}\right)=\eta_{it+1}b_{t+1}-\frac{b_{t+1}^{2}}{2}\quad;\eta_{it}\sim N\left(0,1\right)
\]
$b_{t+1}$ is the cross-sectional standard deviation of consumption
growth. It is specified so that people suffer a high cross-sectional
variance of consumption growth on dates of a low market return $R_{t+1}$.
\[
b_{t+1}=\sigma\left[\ln\left(\frac{C_{it+1}}{C_{it}}\right)\left|R_{t+1}\right.\right]=\sqrt{\frac{2}{\alpha\left(\alpha+1\right)}}\sqrt{\delta-\ln R_{t+1}}
\]
The general model is,
\[
b_{t+1}=\sqrt{\frac{2}{\alpha\left(\alpha+1\right)}}\sqrt{\ln m_{t+1}+\delta+\alpha\ln\frac{C_{t+1}}{C_{t}}}\quad;p_{t}=E_{t}\left[m_{t+1}x_{t+1}\right]\;\forall x_{t+1}\in\underline{X}\equiv\left\{ \text{Set of Payoffs}\right\} 
\]
\[
\ln\left(\frac{\nu_{it+1}}{\nu_{it}}\right)=\eta_{it+1}b_{t+1}-\frac{b_{t+1}^{2}}{2}\quad;C_{it+1}=\nu_{it+1}C_{it}
\]
Using this it is easily shown that, 
\[
1=E_{t}\left[e^{-\delta}\left(\frac{C_{it+1}}{C_{it}}\right)^{-\alpha}R_{t+1}\right]
\]
The excess return can be written as,
\[
0=E_{t}\left[\left(\frac{C_{it+1}}{C_{it}}\right)^{-\alpha}R_{t+1}^{e}\right]
\]
Now aggregating across all consumers by summing over $i$, $E_{N}=\frac{1}{N}\sum_{i=1}^{N}$
and assuming that cross-sectional variation of consumption growth
is log-normally distributed gives,
\[
0=E_{t}\left[E_{N}\left(\left(\frac{C_{it+1}}{C_{it}}\right)^{-\alpha}\right)R_{t+1}^{e}\right]
\]
\[
0=E_{t}\left[\left(e^{-\alpha E_{N}\left[\Delta c_{it+1}\right]+\frac{\alpha^{2}}{2}\sigma_{N}^{2}\left[\Delta c_{it+1}\right]}\right)R_{t+1}^{e}\right]
\]
\end{doublespace}

\end{document}